\documentstyle[12pt,aaspp4]{article}
\input psfig
\def\arcsec {$^{\prime \prime}$}

\def\degr   {$^{\circ}$}
\def\etal   {{\it et~al.\/}}
\def\HI     {H{\sc I}}
\def\HII    {H{\sc~{II}}}
\def\kms    {~km~s$^{-1}$}
\def\mo     {{$M_{\odot}$}}

\begin{document}

\title{Signatures of the Youngest Starbursts:
Optically-thick Thermal Bremsstrahlung Radio Sources in Henize 2-10}
\author{Henry A. Kobulnicky\footnote{Hubble Fellow}  }
\affil{University of California, Santa Cruz \\ Santa Cruz CA, 95064 \\
Electronic Mail: chip@ucolick.org}
\authoremail{chip@ucolick.org}
\author{Kelsey E. Johnson}
\affil{JILA \\ University of Colorado \\ Boulder CO 80309-0440 \\
Electronic Mail: kjohnson@casa.colorado.edu}

\centerline{\it Accepted for Publication in the Astrophysical Journal}

\begin{abstract} 

VLA radio continuum imaging reveals compact ($<$8 pc) $\sim$1 mJy radio
sources in the central 5\arcsec\ starburst region of the blue compact
galaxy Henize 2-10.  While the global radio continuum spectrum is a
power law ($S_\nu \propto \nu^{\alpha}$) indicative of nonthermal
processes ($\alpha\simeq-0.5$), the radio sources have positive
($\alpha>0.0$) spectral indicies suggesting an optically thick thermal
bremsstrahlung origin.  We model the luminosities and spectral energy
distributions of these radio knots, finding that they are consistent
with unusually dense \HII\ regions having electron densities,
$1500~cm^{-3}<n_e<5000~cm^{-3}$, and sizes $3~pc~<R<8~pc$.  While the
high inferred densities are typical of ultracompact \HII\ regions in
the Galaxy, such high optical depth ($\tau=0.4-3.0$ at 5 GHz) at
frequencies as high as 5 GHz is unusual on pc scales in galaxies.
Since these H~II regions are not visible in optical images, we propose
that the radio data preferentially reveal the youngest, densest, and
most highly obscured starforming events.  Energy considerations imply
that each of the five \HII\ regions contains $\sim$750 O7V equivalent
stars, greater than the number found in 30 Doradus in the LMC.  The
high densities imply an over-pressure compared to the typical
interstellar medium so that such objects must be short-lived ($<0.5$
Myr expansion timescales).  We conclude that the radio continuum maps
reveal the very young ($<0.5$ Myr) precursors of ``super starclusters''
or ``proto globular clusters'' which are prominent at optical and UV
wavelengths in He~2-10 and elsewhere.  The fraction of O stars in these
ultra-dense \HII\ regions is 15\% of the total inferred O star
population in He 2-10.  This body of work leads us to propose that
massive extragalactic star clusters with ages $<10^6$ yr, the possible
precursors to globular clusters, may be most easily identified by
finding compact radio sources with optically-thick thermal
bremsstrahlung spectral signatures.

\end{abstract}

\keywords{Galaxies: individual: Henize 2-10 --- 
galaxies: ISM --- galaxies: star clusters --- galaxies: starburst --- HII Regions
 --- stars: Wolf-Rayet}

\section{Introduction}

The blue compact galaxy Henize 2-10 has been extensively studied as the
prototype of starbursting galaxies containing large populations of
Wolf-Rayet stars (e.g., Conti 1991).  Morphological studies reveal at
least three 
distinct starforming regions, termed A, B, and C as show
recently in the images of M\'endez \etal\ (1999).  Each region is
composed of numerous $<$ 10 pc sized OB star clusters (Conti \& Vacca
1994; Johnson \etal\ 1999).  Optical and 21-cm radial velocities of
$v_\odot=$860-870 \kms\ (RC3; Allen, Wright, \& Goss 1976) suggest a
distance between 6 Mpc (Johansson 1987) and  14 Mpc (Allen \etal\ 1976)
depending on the choice of $H_0$ and Virgocentric inflow models.  Thus,
the apparent B-band magnitudes of 12.4 (RC3) to 11.8 (Johansson
\etal\ 1987) result in absolute magnitudes ranging from $M_{B}=-16.5$
(dwarf) to $M_{B}=-18.9~$($\sim$0.6 $L_*$).\footnote{ For
$M_{*B}=-19.4$ as found by Marzke \etal\ 1998.} Assuming the
oft-adopted distance\footnote{The current distance uncertainty
propagates into large uncertainties on other physical parameters
including the star formation rate and young star population.  Further
work on He~2-10 would be greatly aided by a new distance determination
based on methods other than the radial velocity.} of 9~Mpc, the optical
diameter of 60\arcsec\ (RC3) corresponds to 2.8 kpc, making He~2-10 a
comparatively compact galaxy.  The \HI\ rotation curve is solid body in
nature, and single-peaked, with a velocity dispersion ($\sigma_v\equiv
FWHM/2.35$) of 68 \kms\ (Allen \etal\ 1976; Kobulnicky \etal\ 1995).
These properties make He~2-10 similar to some of the compact
star-forming galaxies seen at intermediate redshifts (Koo \etal\ 1995;
Guzman \etal\ 1996, 1998) and in the Hubble Deep Field (Phillips
\etal\ 1997).

He~2-10 contains several $\times 10^8$ \mo\ of molecular gas (Baas,
Israel \& Koornneef 1994; Kobulnicky \etal\ 1995; Meier \& Turner 1999)
and shows evidence for large amounts of dust obscuration (Philips
\etal\ 1984; Beck, Kelly, \& Lacy 1997).  This high dust and molecular
content is consistent with the strong CO emission and with with recent
chemical abundance measurements indicating a metallicity of nearly the
solar value (Kobulnicky, Kennicutt, \& Pizagno 1995).  H$\alpha$ images
of Henize 2-10 show multiple ionized shells and filaments (Beck \& Kovo
1999; M\'endez \etal\ 1999) typical of actively starforming systems
which may be undergoing starburst-driven outflows (e.g. Marlowe
\etal\ 1995; Martin 1998).  Single-dish radio continuum measurements
over a range of frequencies show a dominant nonthermal component,
suggesting the presence significant supernova activity (Allen
\etal\ 1976).

In this paper we present a spatially-resolved radio continuum study of
He~2-10 using multi-configuration data from the Very Large Array
(VLA)\footnote{The National Radio Astronomy Observatory is operated by
Associated Universities, Inc., under cooperative agreement with the
National Science Foundation.}.  We consider the overall radio
continuum energy distribution in light of other recent work on this
galaxy, and we analyze high-resolution (0.5\arcsec) radio maps of the
central starburst in conjunction with HST imaging of Johnson \etal\
(1999).

\section{VLA Radio Continuum Observations}

Aperture synthesis radio continuum observations were carried out with
the NRAO Very Large Array  using a variety of frequencies and array
configurations from 1984 December through 1996 January.  Table~1
summarizes dates, configurations, frequencies, and durations of
observations.  The first four datasets in Table~1 are part of a program
to measure the total radio flux at coarse (6\arcsec) resolution using
``scaled--arrays'' which provide similar UV coverage at 92~cm, 21 cm, 6
cm, and 2~cm (A, B, C, and D configurations).  The other observations
at 2~cm, 3.6~cm, and 6~cm were obtained to study the radio emitting
starburst regions at high angular resolution.

The data were calibrated and mapped within NRAO's
Astronomical Image Processing Software (AIPS).  A nearby point
source, in most cases 0834-201, was observed every 10---15 minutes to
calibrate the relative phases of the antennas.  The radio flux density
standard 3C286 (1328+207) was observed twice per observing session in
order to calibrate the absolute intensities.  Uncertainties on the
absolute fluxes are due to the intrinsic variability of 3C286 ($\sim$1\% ;
Perley 1998) and the magnitude of the closure errors on the secondary
calibrator used to bootstrap the sensitivity solution from the primary
flux calibrator.  These errors varied with each dataset.  We adopt a
median of $\sim$5\% on the zero point of the flux calibration
from epoch to epoch.

\section{Radio Continuum Analysis}
\subsection{Global Radio Spectral Properties}

In order to estimate a total radio flux density at each frequency, we
mapped the 92~cm, 20 cm, 6~cm, and 2~cm low resolution datasets using
only antenna pairs with the shortest baselines by applying a 5
kilo--$\lambda$ Gaussian UV-taper in the AIPS task IMAGR.  These
parameters provide the greatest sensitivity to large extended
structures by weighting short UV baselines most heavily.
The resulting synthesized beam has dimensions
$\sim$26\arcsec$\times$ 24\arcsec.  The typical RMS noise in these maps
is 75 $\mu$Jy beam$^{-1}$ but much larger (280 mJy beam$^{-1}$) at
90~cm due to lower overall sensitivity and confusion from other
background sources in the 130\arcsec\ FWHM primary beam.
Extensive mapping and deconvolution (CLEANing) was necessary to remove
the sidelobe structure from other stronger sources within the primary
beam at 90~cm.  Table~2 summarizes the total radio continuum flux density of
Henize~2-10 at each wavelength, as measured from the low resolution
maps.  

For comparison with the new results, Table~2 lists the single-dish radio
continuum measurements tabulated in Allen \etal\ (1976) over a similar range in
frequency.  We also include the Owens Valley millimeter-wave continuum
measurements at 96 GHz and 230 GHz from Meier \& Turner (1999).
Figure~1 displays these results graphically in a plot of integrated
flux density versus frequency.  In all cases, the VLA flux density is
less than, but within the errors of, the single dish result at the same
(interpolated) wavelength.  This systematic offset is expected since
aperture synthesis telescopes like the VLA lack the short UV spacings
required for a total flux measurement.  Since aperture synthesis arrays
act as spatial filters, the data are insensitive to smooth structures
larger than an angular scale set by the separation of the shortest
antenna pairs.  For the low-resolution configurations and wavelength
combinations used here, structures larger than about 90\arcsec\ will be
invisible to the VLA.  The ratio of flux detected with the
interferometer to single dish results (60\% - 70\%) is comparable to
the ratios for other local starbursts observed with both types of
telescopes (e.g., NGC 5253--Turner, Ho \& Beck 1998).  The largest
discrepancy for He~2-10 occurs at 90~cm.  We interpret this difference
as the result of additional contributions by confusing sources in the
large beam and sidelobes of the single-dish telescopes at low
frequencies.  The total VLA flux at 6~cm and 3.6~cm, 42.6 mJy and 24.3
mJy, respectively, may also be compared to the measurements of M\'endez
\etal\ (1999) using the Australia Telescope Compact Array (ATCA) at
comparable wavelengths.  M\'endez \etal\ measure 30.3 mJy and 21.4 mJy
respectively, 71\% and 88\% of the VLA flux.  This difference may be
explained as the inclusion of shorter projected UV spacings in the 6~cm
C-configuration VLA dataset where baselines as short as 1 kilo-lambda
are present compared to the ATCA dataset of M\'endez \etal.

In Figure~1 we show a best fit power law of the form $S_\nu\propto
\nu^{\alpha}$, where $\alpha=-0.54$ for the VLA data, including the 96
GHz point from Meier \& Turner (1999).  This value is typical of
star-forming galaxies like M~82  
(Klein, Wielebinski, \&
Morsi 1988), NGC~3448 (Noreau \& Kronberg 1987), and H~II galaxies
(Deeg \etal\ 1993) which are generally dominated by nonthermal
(synchrotron) emission with $-0.8 < \alpha < -0.4$).  The 3.6~cm point lies
below a best fit power law to the other data, presumably because the
array configuration combination is not perfectly scaled to match the UV
coverage of the other datasets.  The 3.6~cm dataset lacks the shortest
UV spacings present in the other observations, and thus we do not
consider this dataset in the further analysis of the integrated radio
spectrum.

The overall radio spectral index is highly nonthermal, consistent with
synchrotron radiation produced in supernova explosions.  Our best-fit
value of $\alpha=-0.54\pm0.04$ is slightly flatter than the value
$\alpha=-0.60\pm0.15$ found by Allen \etal\ (1976) from single dish
data over the range 0.4 to 10.6 GHz, and $\alpha=-0.59\pm0.05$ by
M\'endez \etal\ (1999) from two high frequency points (4.8 GHz and 8.6
GHz).  Removing the 90~cm (0.325 GHz) point from the fit yields
$\alpha=-0.58\pm0.05$, more consistent with the other estimates.  The
overall spectrum suggests a combination of thermal and nonthermal
processes, with thermal processes dominating at frequencies of 90~GHz
and above.

He~2-10 shows a low-frequency turnover in its radio spectrum similar to
several of the Blue Compact Galaxies observed by Deeg \etal\ (1993).
Deeg \etal\ considered several physical mechanisms to explain this
low-frequency turnover.  Free-free absorption of the predominantly
synchrotron spectrum by thermal electrons in the halo of He~2-10 could
explain this result by modifying the predominantly nonthermal spectrum
at long wavelengths.

\subsection{High Resolution Radio Morphology and Spectra}
\subsubsection{Overall Morphology}

In order to investigate the radio morphology of Henize~2-10 on small
angular scales, we selected the 3.6~cm (8.4 GHz) data from 1996 May 14
which provides the best combination of angular resolution, sensitivity,
and UV coverage.  These data have roughly 2--3 times the angular
resolution of the ATCA dataset presented in M\'endez \etal\ (1999) but is
less suitable for measuring total fluxes or extended structures due to
a relative lack of short UV spacings.  Figure~2 displays a 3.6~cm (8.4
GHz) contour map made with a UVTAPER of 500 kilo-wavelengths.  The  RMS
noise level is 35 mJy beam$^{-1}$.  The synthesized beamsize of
0.57\arcsec$\times$0.70\arcsec\ corresponds to a linear size of
25$\times$30 pc for the assumed distance of 9 Mpc.  Contours are -5,
-4, -3, 3, 4, 5, 6, 8, 10 12, 14, 16, 18, 20, and 22 times the RMS
noise level.  Figure~2 also shows a Hubble Space Telescope WFPC2 image
obtained through the F555W filter (Johnson \etal\ 1999) presented as
a logarithmic greyscale.  The absolute coordinates of the radio map are
accurate to 0.05\arcsec\, while the HST image are uncertain by
$\sim$0.5\arcsec, limited by the positional uncertainties of the guide
stars from the Guide Star Catalog (Lasker \etal\ 1990).  This leads to
a relative positional uncertainty between the radio and optical images
of the same amount, 0.5\arcsec.

Figure~2 shows several radio ``knots'' aligned in an east-west
orientation.  The overall radio morphology is reminiscent of that in
M~82 (Golla, Allen, \& Kronberg 1996) and NGC~3448 (Noreau \& Kronberg
1987) observed at similar linear resolution.  We identify five radio
knots which we label Knots 1 through 5 from west to east. Our Knot 4 corresponds to RC1 in the radio maps of M\'endez
\etal\ (1999) while our Knots 1 and 2 together comprise their RC2. 
Figure~2 does not show the spur of radio emission extending eastward from RC1
in the M\'endez \etal\ maps, even though the data presented in Figure~2
has the same frequency and higher sensitivity (RMS=35 $\mu$Jy
beam$^{-1}$) than the ATCA data (RMS=150 $\mu$Jy beam$^{-1}$).  These
maps do, however, show the spurs of radio emission extending
northward from RC1 and RC2 (our Knots 4 and Knots 1\&2), lending
credence that these are real features somehow associated with the
starburst events. The signal-to-noise is not sufficient in these
regions to obtain a spectral index (see below) but their relative
strength at 6~cm compared to 3.6~cm suggests a nonthermal origin.
The proximity of these features to the nuclear star clusters
is consistent with a scenario wherein wind-driven and supernovae shells 
sweep up and compress the magnetic field along with high-energy
electrons to produce regions of enhanced synchrotron emission.

The radio knots contribute 4.4 mJy of the 24.3 mJy total at 3.6~cm.
Each knot contributes just under 1 mJy to this total.  They have finite
deconvolved sizes less than 0.1\arcsec--0.5\arcsec\ corresponding to
linear sizes of 3-20 pc.   
Diffuse emission in Figure~2 which comprises the
majority of the radio flux is nearly aligned with the optical major
axis of the galaxy at 168\degr\ (Corbin, Korista, \& Vacca 1996).  This
high resolution map contains nearly all of the flux seen in lower
resolution maps made with the same detaset, but due to the lack of
short spacing in the BnA configuration, 24.3 mJy should be regarded as
a lower limit to the total flux at 3.6~cm as seen with a single dish.

Although there is a relative uncertainty of
0.5\arcsec\ between the radio and optical images, we can say that there
is no obvious correlation between the morphology 6-8 optical/UV star
clusters (distributed in a chevron configuration) and the radio
sources.  Given any possible 0.5\arcsec\ translation of the optical
image, no more than 2 of the radio sources would have a corresponding
optical/UV counterpart.  Given the high extinction in the core of this
galaxy, it is not surprising that the optical and radio morphologies
are so different.  The ground-based JHK infrared images of
Beck \etal\ (1997) have somewhat lower (0.9\arcsec) resolution and do not
show prominent features which could correspond to
the radio knots.  

\subsubsection{Radio Continuum Spectral Index}

In an effort to study the spectral
properties of these compact structures, we mapped the high resolution 2~cm
and 6~cm datasets, taking care to use only matched UV data from array
configurations A and B.  The only observations with sufficient quality (high
resolution and good phase stability) were the A configuration 6--cm
data from 1995 and B configuration 2--cm data from 1996.  Because of
the snapshot nature of the observations, the UV coverage is slightly
different for each dataset, but the overall beamshape is nearly
identical.  Mapped with a Gaussian taper of $\sim$500 k$\lambda$, the
UV data produced a clean beam size of 0.82\arcsec$\times$0.40\arcsec,
highly elongated in the north-south direction.  The UV coverage of the
3.6~cm data does not match the other datasets, so we do not consider it
for comparison here.

The 2~cm and 6~cm maps appear in contour form in Figs.~3a--3b overlaid
on the HST WFPC2 F555W image of the central region of He~2-10 (Johnson
\etal\ 1999).  Because the data are mapped to obtain the highest
possible angular resolution, only the high surface brightness radio
features are visible.  The total flux in the maps is 17.5 mJy at 2~cm and
14.5 mJy at 6 cm, 83\% and 34\% respectively, of the total flux
measured with lower resolution VLA configurations and mapping
procedures.  Table~2 records the total flux in these maps at each
frequency for each dataset.  From these maps it is clear that at least
6 distinct radio components make up the central region of Henize 2-10.
We performed multiple Gaussian fitting with the AIPS task IMFIT to
characterize the positions and relative strengths of these sources.
 We list the positions and fluxes of these knots in Table 3 based on
elliptical Gaussian fits to maps made with the 2~cm and 6~cm datasets
as described below.  Uncertainties on the peak and integrated flux
densities are dominated by limitations in defining the background flux
levels around each knot.  We estimate the total flux uncertainties at
20\% based on multiple attempts to fit each component.  Only five of
the sources could be successfully fit and deconvolved due to the
irregular background and limited angular resolution.  Each of the
sources has an integrated flux near 1 mJy.  Table ~3 
shows that the compact radio sources comprise 27\% (2~cm)
and 9\% (6~cm) of the total radio emission in Henize 2-10 as measured
with the VLA.  For the assumed distance of 9 Mpc, 1 mJy corresponds to
a luminosity of 9.2$\times10^{25}$ erg s$^{-1}$ Hz$^{-1}$ or
9.2$\times10^{18}$ W Hz$^{-1}$.  This is similar to the luminosity of
young (few yr) radio supernovae reported by Weiler \etal\ (1989) in a
variety of local galaxies.  However, we find the same radio sources
present in the datasets from 1984.  Although
the 6~cm radio data from 1984 and 2~cm data from 1985 are of lower
quality (about half the sensitivity), the strengths of the radio
components are comparable, within the errors, of the more recent data.
Since radio supernovae have typical decay times of a few years (Weiler
\etal\ 1986), it is unlikely that these sources still present ten years
later in 1996 can be such transient events.

>From Figure~3 it is clear that the compact radio sources are more
luminous at shorter wavelengths, and thus have an inverted 
(positive $\alpha$) radio spectral index.  Figure~4 shows a map of the
spectral index, $\alpha^6_2$ (where $S_\nu\propto\nu^\alpha$),
constructed from the 2 and 6~cm maps which were first blanked in
regions with signal-to-noise $<$4:1.  The spectral index, shown in
greyscale, varies from $\alpha=-0.5$ (mostly nonthermal) to
$\alpha=-0.1$ (pure optically thin thermal bremsstrahlung) to
$\alpha=1.0$ (optically thick thermal bremsstrahlung).  Contours show
the 2 cm radio continuum from Figure~3a.  Most starburst galaxies show
radio spectra dominated either by synchrotron emission ($-1.2 < \alpha
< -0.4$), or thermal bremsstrahlung ($\alpha=-0.1$), or some
combination (e.g., Deeg \etal\ 1993).  Since radio supernovae typically
have negative spectral indicies after several hundred days (Weiler
\etal\ 1986), the  positive spectral index of these sources is
additional evidence that they are not young supernovae.

Might these radio sources be supernova remnants which have much longer
lifetimes than the individual supernovae themselves?  The luminosities
of typical SNRs in the Milky Way are several orders of magnitude less
luminous, ranging from 8$\times10^{21}$ erg s$^{-1}$ Hz$^{-1}$ for 1000
yr old remnants, to 7$\times10^{24}$ erg s$^{-1}$ Hz$^{-1}$ for Cas A
(330 yr).  Each of the five $\sim$1 mJy sources is thus a factor
of 10 more luminous than Cas A.  At the assumed distance of He~2-10, a
fairly radio-luminous young supernova remnant like Cas A would
be only a 2$\sigma$ detection in the current data.  

M~82 also contains multiple supernova remnants making it a good object
for comparison with He 2-10. With a distance of $\sim$3.63 Mpc
(Freedman \etal\ 1994), it can be studied at higher spatial
resolution.  Radio continuum maps presented in Allen \& Kronberg (1998)
show 26 compact sources with linear diameters of 3.5 pc (0.2\arcsec)
and 2~cm spectral luminosities of 2.9$\times10^{24}~erg~s^{-1}~Hz^{-1}$
to 7.3$\times10^{25}~erg~s^{-1}~Hz^{-1}$ (fluxes of 0.2 mJy to 5 mJy).
The five radio sources in He~2-10 fall at the upper end of this
luminosity range.  Of the 26 sources in M~82, 22 show distinctly
nonthermal spectral indicies and are consistent with supernova remnants.
Their radio spectra often show turnovers, but only at low frequencies
($\nu< 1$ GHz), due to free-free absorption in the surrounding ionized
gas.  Only one source has a thermal spectrum typical of H~II regions.
Three sources remain unidentified.  By comparison, the five radio
sources in He 2-10 have large luminosities and inverted spectral
indicies at frequencies as high as $\nu>5$ GHz, suggesting that none of
them are individual SNRs.  Although the nonthermal nature of the
global radio spectral index in He 2-10 clearly indicates that several thousand
supernovae contribute to the observed synchrotron emission, the
individual events are probably too old and faint to be seen except
collectively. 

Compact H~II regions are strong sources of free-free emission and
represent most plausible origin of the compact radio knots.
Figures~5a \& 5b show the radio continuum contours at 2~cm and 6~cm
superimposed on a continuum-free HST narrowband $H\alpha$ image.  The
relative alignment of the two images is accurate to
$\sim$0.5\arcsec\ RMS.  There is no obvious correlation between the
radio and $H\alpha$ features, although if the HST image is shifted
$\sim$1\arcsec\ to the east, several of the radio maxima would more
closely correspond to the regions of highest emission line surface
brightness.  Unfortunately it is not possible to unambiguously identify
the radio features with $H\alpha$ counterparts.  
If the radio knots are due to \HII\ regions, then the lack
of correspondence with the $H\alpha$ images suggests a large
amount of extinction.  Narrowband
near-infrared imaging in the Bracket $\gamma$ line
at high spatial resolution (0.1\arcsec) could help
confirm the \HII\ region nature of the sources.

Flat or positive spectral indicies like the ones observed in He~2-10 are
common for ultra-compact H~II regions in the Galaxy on sub pc scales (e.g,
see the compilation of Wood \& Churchwell 1989a for ultra-compact H~II
regions), but are very unusual on larger pc-sized scales in other
galaxies.  Turner, Ho, \& Beck (1998) report a radio source in
the nucleus of NGC~5253 which has about the same luminosity as the each
of the 5 sources we find in He~2-10.  It has a flat spectral
index consistent with mostly thermal emission.  Allen \& Kronberg
(1998) report a radio source, 42.21+590, with a similar luminosity
(6.9$\times10^{25}~erg~s^{-1}~Hz^{-1}$) in M~82 which they identify as
an H~II region on the basis of its flat ($\alpha=-0.1$) radio spectral
index.  All five radio sources in He~2-10 show radio spectral indicies
greater than $\alpha=0.0$, suggesting that their spectral energy
distributions have been modified by free-free absorption.  

In Figure~6
we plot the 2~cm (14.9 GHz) and 6~cm (4.8 GHz) fluxes of each
source to show the rising nature of the radio spectrum toward higher
frequencies.  Since the absorption coefficient for free-free
absorption, $\kappa_\nu$, is proportional to $\nu^{-2}$, there is a
turnover frequency $\nu_t$, where a plasma becomes optically thick to
radiowave frequencies and the spectral index changes from $\alpha=-0.1$
to positive values, approaching the blackbody limit of $\alpha=2$ as
$\tau\rightarrow\infty$.  The brightness temperature is given by

\begin{equation}
T_B(K)={{F\lambda^2}\over{2k\Omega}},
\end{equation}

\noindent where $k$ is Boltzman's constant,
and $\Omega$ is the beamsize.  In useful units, this becomes,

\begin{equation}
T_B(K)=1541.5\times4ln2{{F(Jy)\lambda^2(cm)}\over{\pi{a}('')b('')}}
\end{equation}

\noindent where $F$ is the flux of the source, $\lambda$ the wavelength
of observation, and $a$ and $b$ are the major and minor FWHM of the
synthesized beam in arcseconds.  In the absence of detailed knowledge about the
structure of each knot, we assume that the emission fills the beam
uniformly.  This assumption means that an estimate of the brightness
temperature is necessarily a lower limit on the actual value.  We find
brightness temperatures between 15~K and 28~K at 2~cm, increasing to 81
to 164~K at 6~cm as summarized  in Table~3.  We estimate a {\it lower limit}
on the radio optical depth by comparing the brightness temperature,
$T_B$, of the sources to the electron temperature of the gas, $T_e$,
using

\begin{equation}
T_B=T_e(1-e^{-\tau})
\end{equation}

\noindent Adopting $T_e$=6000~K based on the spectroscopy of Kobulnicky
\etal\ (1999) and the brightness temperatures in Table~3, we find
optical depths of $\tau\geq$0.01 --- 0.02 at 6~cm.  Using Mezger \&
Henderson (1967), the emission measure $EM={\int}n^2_edl$ is

\begin{equation}
EM(cm^{-6}~pc)= 0.083\Big({{T_e}\over{(K)}}\Big)^{-1.35}
\Big({{\nu}\over{(GHz)}}\Big)^{-2.1}  \tau.
\end{equation}

\noindent These knots yield emission measures in excess of 
$5 \times 10^4$ $cm^6~pc$.  

In order to explore the nature of the compact sources in more detail
without needing to make assumptions about the size of the emitting
region, we constructed model H~II regions (spherical, uniform electron
density and electron temperature) of radius, $R$, electron temperature,
$T_e$, an electron density, $n_e$.  We computed the emergent radio
spectrum considering only thermal Bremsstrahlung emission and
self-absorption processes.  In Figure~6 we show the resulting spectral
energy distributions for a range of models. The luminosities and
spectral indicies of these compact sources are well-fit by model H~II
regions with a radius of 3---8 pc (0.06\arcsec--0.17\arcsec), a mean
electron temperature of 6000~K, and electron densities of 1500-5000
$cm^{-3}$ at a distance of 9 Mpc.  These 
sizes imply that the emitting regions fill only 4\% to 40\% of the VLA
synthesized beam.   The optical depths at 4.8 GHz range
between $\tau=$0.3 and $\tau=$2.5 for the different models.  Emission
measures range between $10^7$ $cm^6~pc$ and $10^8$ $cm^6~pc$.  
The required densities are
consistent with $500<n_e<11000$ measured from optical spectroscopy of
the nuclear regions (Hutsemekers \& Surdej 1984).
The model \HII\ regions have optical depths $\tau_{5~GHz}\approx0.4-3$ 
and $\tau_{15~GHz}\approx0.04-0.3$.

We plot the spectral energy distribution of several model
bremsstrahlung sources in Figure~6 to show that this range of electron
densities and radii can reproduce the observed luminosities and
spectral energy distributions.  Such high electron densities are
required to produce the large optical depth at frequencies lower than 5
GHz.  The source size is constrained by the requirement to match the
total luminosity of the knots and to be consistent with the 
upper limits on angular sizes measured
in the VLA maps.  Based on the large optical depths
due to free-free absorption at frequencies below 5 GHz, we predict that
these radio knots will be extremely faint at longer wavelengths
($<100~\mu{Jy}$ at 1.4 GHz) where the optical depth should
exceed $\tau_{1.4~GHz}>20$ for some knots.

\section{Energy Budget and Ionizing Stars}

\subsection{He~2-10 as a Whole} 

The total H$\alpha$ flux of He~2-10 reported by Beck \& Kovo (1999) is
1.35$\times10^{-10}~erg~s^{-1}~cm^{-2}$ which corresponds to a
luminosity of 1.2$\times10^{42}~erg~s^{-1}$ at 9 Mpc.  This is about an
order of magnitude higher than M\'endez \etal\ (1999) who report an
H$\alpha$ flux of $1.82\times10^{-11}~erg~s^{-1}~cm^{-2}$.  
Published H$\alpha$ slit spectroscopy (Vacca \& Conti 1992) 
and new HST H$\alpha$ imaging (Johnson \etal\ 1999) is more consistent with the latter
value, so we adopt the Mend\'ez \etal\ measurement 
which implies an H$\alpha$ luminosity of
1.6$\times10^{41}~erg~s^{-1}$ at 9 Mpc.  For a case B ionization
bounded nebulae in which all of the H-ionizing photons are re-emitted
as Balmer series photons, the H$\alpha$ luminosity indicates a Lyman
continuum photon production rate\footnote{ We use, $Q_{Lyc}~(s^{-1})=
{{1}\over{h\nu_{H\alpha}}}{{\alpha_B(10,000~K)}\over{\alpha_{H\alpha}^{eff}(10,0
00~K)}}
L_{H\alpha}=7.3\times10^{11}~L_{H\alpha}~(erg~s^{-1}$).  This should be
regarded as a lower limit on $Q_{Lyc}$ since the galaxies may be
translucent to Lyman continuum photons.} of
$Q_{Lyc}=1.1\times10^{53}~s^{-1}$.
The thermal radio luminosity
can be used to obtain a second estimate of $Q_{Lyc}$ following Condon
(1992),

\begin{equation}
\Big({{Q_{Lyc}}\over{s^{-1}}}\Big)
\geq6.3\times10^{52}\Big({{T_e}\over{10^4~K}}\Big)^{-0.45} 
\Big({{\nu}\over{GHz}}\Big)^{0.1}
\Big({{L_{thermal}}\over{10^{27}~erg~s^{-1}~Hz^{-1}}}\Big)
\end{equation}

\noindent Adopting the total measured 2~cm (15 GHz) flux of 21.1 mJy
from Table~2 ($L_{2~cm}= 1.9\times10^{27}~erg~s^{-1}~Hz^{-1}$), we
estimate ${Q_{Lyc}}=2.0\times10^{53}~s^{-1}$, assuming that all of the
2~cm flux is due to free-free processes and adopting a mean
electron temperature of 6000~K for He~2-10 (Kobulnicky, Kennicutt, \& Pizagno
1999).  Corrected by a factor of 1.9 (0.7 mag) for
extinction\footnote{Although there are clearly large extinction
variations within He~2-10, we estimate the global extinction in He~2-10
from the H$\alpha$/H$\beta$ line ratios of Kobulnicky, Kennicutt, \&
Pizagno (1999) which suggest a logarithmic extinction parameter
c(H$\beta$) =0.54 corresponding to $A_V=2.1c(H\beta)$=1.1 mag and
$A_{H\alpha}$=0.7 mag for a Seaton (1979) reddening law.}  the M\'endez
\etal\ measurement yields ${Q_{Lyc}}=2.1\times10^{53}~s^{-1}$,
consistent with the radio estimate of
${Q_{Lyc}}=2.0\times10^{53}~s^{-1}$.  Assuming that a typical O star
(type O7V) produces $Q_{Lyc}=1.0\times10^{49}~s^{-1}$ (Vacca 1994;
Vacca, Garmany, \& Shull 1996), then 20,000 such O stars are required
to power the observed emission.  This estimate is greater than the
number of O5 stars derived by Vacca \& Conti (1992) on the basis of
nebular spectroscopy of region A, and is somewhat lower than the 31,000
O stars estimated by Conti \& Vacca (1994) from UV imaging.  However,
the exact number of stars is highly dependent on the assumed age of the
starburst, so the different methods are not necessarily in
disagreement.

Adopting the conversion from H$\alpha$ luminosity to star formation rate
(SFR) of Kennicutt (1983) 

\begin{equation}
SFR (M_\odot~yr^{-1}) = 8.9\times10^{-42}~{L_{H\alpha}}~(erg~s^{-1}),
\end{equation}
we find a star formation rate of 1.4 $M_\odot~yr^{-1}$.
For the 11.9 Mpc distance adopted by M\'endez \etal\ (1999)
this would become 2.5 $M_\odot~yr^{-1}$.  

\subsection{The Dense H~II Regions}

The energy requirements for each of the compact radio-emitting knots
can be computed in a similar manner to the estimate for the entire
galaxy, using Equation~5 above.  Using the 2~cm luminosities in
Table~3, we find that each knot must be powered by 5--10$\times10^{51}$
Lyman continuum photons $s^{-1}$.  Assuming that each O7V star produces
$Q_{Lyc}=1.0\times10^{49}~s^{-1}$ (Vacca 1994; Vacca, Garmany, \& Shull
1996), each knot requires between 500 and 1000 such stars as its energy
source.  Thus, each knot contains more O stars than are found in the
entire 30-Doradus nebula in the LMC ($\sim200$, Vacca \etal\ 1995;
$\sim$80, Parker 1992).  Using the Starburst~99 models of Leitherer
\etal\ (1999) with solar metallicity, Salpeter IMF, an upper mass
cutoff of $100~M_\odot$, and a lower mass cutoff of $1~M_\odot$,
starburst knots producing this range of Lyman continuum photons at an
age of 1 Myr would have masses of 0.8--1.6$\times10^5~M_\odot$.  These
masses are consistent with those found for optical knots in He2-10 by
Johnson \etal\ (1999).

Since the star-forming regions implied by these radio features are not
seen even faintly on the HST V or I-band images, their visual
extinctions must be quite large, $A_V>5$.  Such large extinction values
are consistent with the column density of molecular gas observed in
this region (Kobulnicky \etal\ 1995) and the extinction estimates from
the infrared silicate features ($A_V\simeq30$, Phillips \etal\ 1984).
Given that He~2-10 was already known to contain between 4600 (Vacca \&
Conti 1992; a lower limit due to the longslit spectroscopic nature of
the observations) and 31,000 O stars (Conti \& Vacca 1994) the additional
$\sim$4000 O stars contained in these five dense, heavily obscured H~II
regions represent a significant faction of the massive stars in
He~2-10.  We list in Table~3 the H$\alpha$ luminosity of each knot
based on the thermal radio continuum luminosities.  From these numbers
it is easy to predict Bracket $\gamma$ or other emission line
luminosities using recombination coefficients for hydrogen tabulated by
Hummer \& Storey (1987).

If all H~II regions go through an enshrouded phase, it is possible to
estimate the fraction of their lifetimes which these objects spend
buried in their parent molecular clouds.  This fraction is simply the
number of O stars free of the surrounding material (which therefore can
be detected in the optical bands) divided by the number of O stars
which are still behind such high column densities that they are only
detectable at longer wavelengths, such as the radio regime.  In this
case, if we adopt the number of O stars detected in the optical as
31,000 (Conti \& Vacca 1994) in comparison the $\sim$4000 we have
detected in heavily obscured H~II regions, O stars spend approximately
13\% of their lives embedded in optically thick regions.  This is
remarkably consistent with the results of Wood \& Churchwell (1989a)
who find that massive stars in Galactic ultra-compact \HII\ regions
spend approximately 15\% of their lives within their parent molecular
clouds.

\section{Discussion and Conclusions}

Our major result is the discovery of multiple optically thick free-free
sources, probably giant H~II regions, which comprise a substantial
fraction (27\%) of the total radio luminosity at 15 GHz.  The majority
of these sources do not coincide with features visible in the optical
HST images, leading us to conclude that they are heavily obscured by
dust.  The combination of high obscuration and high density inferred
from the free-free optical depth is consistent with very dense H~II
regions such as the one seen in NGC~5253 (Turner, Ho, \& Beck 1998).
Electron densities as high as $10^5$ have been measured in He~2-10 from
the optical spectroscopy of Hutsemekers \& Surdej (1984).  While such
dense, optically-thick, inverted-spectrum H~II regions exist in
abundance around {\it individual} stars in the Galaxy (i.e.,
ultra-compact \HII\ regions; Wood \& Churchwell 1989a) this phenomenon
has not been previously seen on such large spatial and energetic
scales.  Because of their similar densities to ultracompact \HII\
regions in the Galaxy, we might term them ``ultra dense \HII\ regions''
(UD\HII).  Their luminosities suggest that such dense \HII\ regions
harbor the youngest ``super star clusters'' which are
observed to be ubiquitous in He 2-10 and other starburst systems (e.g.,
Conti \& Vacca 1994; O'Connell, Gallagher, \& Hunter 1994; O'Connell
\etal\ 1995; de Marchi \etal\ 1997).  These high densities are 
similar to the densities required on theoretical grounds in the birthplaces
of globular clusters (Elmegreen 1999).

It is not surprising that a fraction ($\sim$4000 out of 5000-31,000) of
the massive stars in He2-10 are hidden by dust, perhaps in their parent
molecular clouds.  After all, massive stars seem to live a significant
fraction of their lives ($\sim15$\%; Wood \& Churchwell 1989b) buried
in their natal material.  What {\it is} surprising, is that so {\it
many} OB stars are located in these massive high-density H~II regions
in He~2-10 compared to other galaxies.  He~2-10 contains 5 such H~II
regions containing $\sim$4000 O7V equivalent stars, while the more
luminous starbursts like M~82 contain one, at most (Allen \& Kronberg
1998).  What makes He~2-10 special?  With electron densities of up to
5000 $cm^{-3}$ compared to the $n_e\simeq$100 $cm^{-3}$ observed in
typical giant extragalactic H~II regions, the five radio-bright H~II
regions in He~2-10 should have an overpressure, compared to the
surrounding ionized gas with similar temperature of,
$\sim5000/100\geq50$.  $P/k\sim10^6$ for the typical
ionized ISM in galaxies whereas $P/K\sim10^7-10^8$ here.
Such high pressure regions should expand rapidly to
re-establish pressure equilibrium and become undetectable on timescales
of few$\times10^5$ years assuming an expansion rate equal to the sound
speed, 10 \kms.  Thus, they do not exist for longer than 10\%-15\% of
the lifetime of a typical O stars unless confined by an additional
external pressure greater than that in the typical ISM.

Perhaps the pressure of the infalling molecular material seen in
millimeter-wave interferometer maps (Kobulnicky \etal\ 1995; Meier \&
Turner 1999) provides the confinement needed to prolong the lifetime of
these giant, dense \HII\ regions.  Or, perhaps other compact
star-forming galaxies, along with nuclear starburst systems, contain
dense, heavily obscured H~II regions like the five in He~2-10, but they
have simply not been identified as such.  One recent radio continuum
study of NGC~2146 provides evidence that such high-density H~II regions
will be found in other galaxies as well (Tarchi \etal\ 1999).  High
resolution 2~cm and 6~cm radio observations of a whole sample of
starburst systems acquired in scaled VLA A and B-array configurations will
be required to know for certain.  To date, such high spatial resolution
data generally exists only for more luminous systems like quasars and
AGN.  We predict that sensitive multi-frequency imaging of normal
starburst systems will show that this dense \HII\ region phenomena is
simply a normal, but short-lived phase in the chronology of a
starburst.  The short lifetime implied for such structures means that
their presence is an unmistakable signature of the youngest ($<1$ Myr)
super starclusters.  Since super starclusters are proposed to be young
versions of globular clusters, we may be seeing the birth of globular
clusters in the nucleus of Henize 2-10.

\acknowledgments 

Special thanks to David E. Hogg for sharing his 1984-1985 epoch VLA
observations and for helpful comments on the manuscript.  We
acknowledge the delightful atmosphere, stimulating conversations and
inspiration from colleagues at IAU Symposium 193 on Wolf-Rayet
Phenomena in Galaxies in Puerto Vallarta, November 1998, especially
Sara Beck, David M\'endez, Cesar Esteban,  David Meier, and Jean
Turner.  It is also our pleasure to thank Peter Conti and Bill Vacca
for continued discussions with us on these subjects.  We thank Uli
Klein for comments on a draft manuscript and for sharing the results of
a similar radio study of NGC~2146 in advance of publication.  
We thank David M\'endez and David Meier for sharing results
and data in advance of publication. H~.A.~K.
thanks Evan Skillman and John Dickey for guidance and encouragement
advice during the early phases of this project.  H.~A.~K.  is grateful
for support from a NASA Graduate Student Researchers Program fellowship
and a Hubble Fellowship \#HF-01094.01-97A awarded by the Space
Telescope Science Institute which is operated by the Association of
Universities for Research in Astronomy, Inc. for NASA under contract
NAS 5-26555.  K~.E.~J. is pleased to acknowledge support for this work
provided by NASA through a Graduate Student Researchers Fellowship and
grant GO-06580.0195A.

\clearpage

\begin{figure} 
\centerline{\psfig{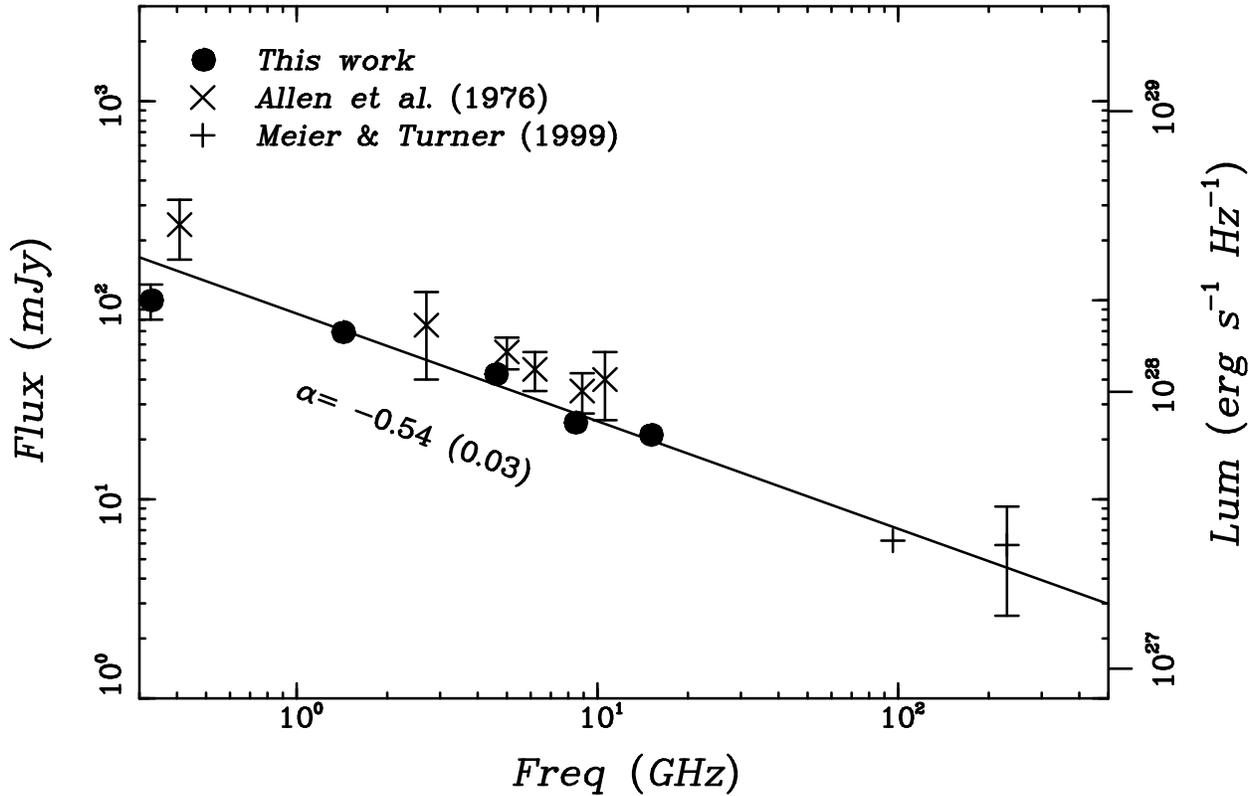}}
\figcaption[f1.ps] {Global radio continuum fluxes of Henize 2-10
as a function of frequency.  Filled circles denote data from
the VLA measurements reported here, and x's denote the single-dish
measurements from Allen \etal\ (1976).  The best fit power law
to the VLA data, including the 96 GHz point
from Meier \& Turner 1999 ($S_\nu\propto\nu^{\alpha}$) has a slope of 
$\alpha=-0.54\pm0.03$.  This slope indicates
that nonthermal processes are the dominant source of
radio luminosity, as is typical of starforming systems 
(e.g., Deeg \etal\ 1993).  \label{radio_spec} } \end{figure}

\begin{figure} 
\figcaption[f2.ps] {Hubble Space Telescope WFPC2 F555W broadband image
of Henize 2-10 (greyscale; Johnson \etal\ 1999) and a VLA 3.6~cm
radio continuum map (contours) with a synthesized beamsize of
0.70\arcsec$\times$0.57\arcsec\ FWHM.  Contours levels are -5, -4, -3,
3, 4, 5, 6, 8, 10 12, 14, 16, 18, 20, and 22 times the RMS noise level
of 35 $\mu$Jy beam$^{-1}$.  The beamshape appears at lower left.   The
greyscale representation is logarithmic, in arbitrary units.   The
absolute coordinates of the radio and HST images are good to
$\leq$0.1\arcsec\ and $\sim$0.5\arcsec\, respectively, so that the
relative positions are uncertain by 0.5\arcsec\ RMS.
Five distinct radio sources are visible.
The lack of a strong correlation between the radio and optical
morphology suggests dust extinction is a significant factor.
\label{F555-3.6best} } \end{figure}

\begin{figure}
Fig~3a---
Hubble Space Telescope WFPC2 F555W broadband image of Henize 2-10
(greyscale) and a VLA B configuration 2~cm radio continuum map
(contours) with a synthesized beamsize of
0.82\arcsec$\times$0.40\arcsec\ FWHM shown at lower left.  Contours
levels are -5, -4, -3, 3, 4, 5, 6, 8, 10 12, 14, 16, 18, 20, and 22
times the approximate RMS noise level of 50 $\mu$Jy beam$^{-1}$. 
Fig~3b---
HST WFPC2 F555W broadband image of Henize 2-10
(greyscale) and a VLA A configuration 6~cm radio continuum map
(contours) with a synthesized beamsize of
0.82\arcsec$\times$0.40\arcsec\ FWHM shown at lower left.  Contours
levels are -5, -4, -3, 3, 4, 5, 6, 8, 10 12, 14, 16, 18, 20, and 22
times the approximate RMS noise level of 50 $\mu$Jy beam$^{-1}$.
The five prominent radio sources are stronger at 2~cm than 6~cm,
characteristic of optically-thick thermal Bremsstrahlung sources.
end{figure}
\clearpage

 \end{figure}

\begin{figure} 
Fig~4---Spectral index map (greyscale) made from the 2~cm and 6~cm
high resolution maps superimposed on the 2~cm contour map
from Figure~3a.  Only the highest surface-brightness
regions appear in these maps.  
The input 2~cm and 6~cm maps were each clipped
at $4\sigma$ so that the regions with
spectral index uncertainties, $\Delta\alpha^6_2$, larger than
0.17 are blanked.  The spectral indicies, $\alpha^6_2$ range from
$\alpha=-0.5$ (mostly nonthermal) to $\alpha=-0.1$ (pure optically thin
thermal bremsstrahlung) to $\alpha=1.0$ (optically thick
thermal bremsstrahlung).  The spectrali and luminosity
characteristics
of these sources are most consistent with
dense ($n_e>1500~cm^{-3}$) \HII\ regions. 
\end{figure}

\begin{figure} 
Fig~5a---H$\alpha$ narrowband image of Johnson \etal\ (1999) with
the 2~cm radio continuum from Figure~3a in contours.  
The $H\alpha$ image is shown as a logarithmic
greyscale in arbitrary units.  There is no clear
correspondence between radio features and $H\alpha$ features, but the
relative alignment of the two images is accurate to only 0.5\arcsec\ RMS.
Shifting the $H\alpha$ image to the east by $\sim$1\arcsec\ would
produce a reasonably good correspondence between the radio 
and emission line maxima, but the different morphologies
suggest large amounts of extinction in the nucleus.  
Infrared emission line imaging could confirm the presence of highly
obscured, young ($<500,000$ yr) \HII\ regions which are seen here only in the radio
continuum.
Fig~5b---H$\alpha$ narrowband
images of Johnson \etal\ (1999) with the
6~cm radio continuum from Figure~3b in contours.
\end{figure}

\begin{figure} 
\end{figure}

\begin{figure} 
\centerline{\psfig{file=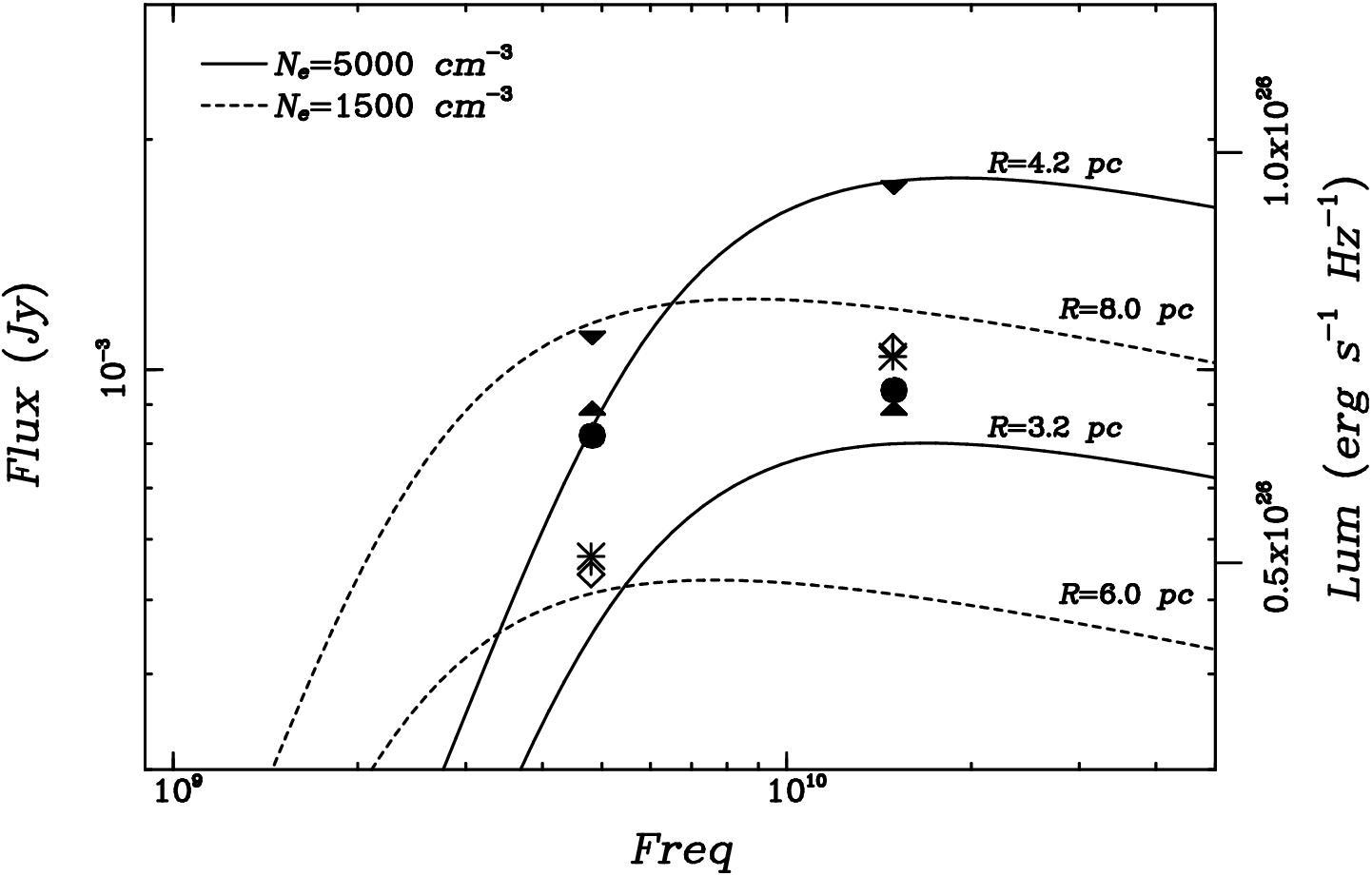,width=6.0in,angle=-90}}
Fig~6---VLA 6cm (4.8~GHz) and 2~cm (14.9 GHz) fluxes and luminosities
for the five radio knots in Henize~2-10.  A different symbol represents
data for each knot listed in Table~3.  Solid and dotted lines represent
thermal bremsstrahlung plasmas modeled as
spheres of radius, $R$, electron temperature $T_e$=6000~K, and mean
electron densities of 1500 $cm^{-3}$ and 5000  $cm^{-3}$ respectively.
Such high densities are required to produce a sufficiently large
free-free opacity needed to explain the rising spectral index between
4.8~GHz and 14.9~GHz.  Thermal bremsstrahlung sources
(i.e., H~II regions) with radii between 3 pc and 8 pc
are consistent with the imaging data and can
reproduce the range of spectral shapes and luminosities of the
radio knots. These high densities imply an overpressure
compared to typical warm ionized medium pressures.
Such \HII\ regions should expand and become undetectable 
in the thermal radio continuum on timescales of 500,000 yr. 
\end{figure}

\end{document}


\begin{deluxetable}{crrrrrl}
\tablewidth{0pt}
\tablecaption{VLA Observational Parameters  \label{Obs} }
\tablehead{
\colhead{Date}   & 
\colhead{Config.}   & 
\colhead{freq.} & 
\colhead{BW} & 
\colhead{$\lambda$} & 
\colhead{Int. Time} &
\colhead{Comments} \\ [.2ex]
\colhead{ } &  
\colhead{ } &  
\colhead{(GHz)} &  
\colhead{(MHz)} &  
\colhead{(cm)} &  
\colhead{(min)} &
\colhead{}
} 
\startdata
Low Resolution Configurations \nl
1993 Dec 20  & D    & 15.0 & 2x50       & 2   & 15 & ok  \nl
1994 Oct 23  & C    & 4.58 & 2x50       & 6   & 15 & ok  \nl
1994 May 14  & BnA  & 1.420 & 20x0.024  & 21 & 240 & ok \nl
1994 Apr 25  & A    & 0.327 & 2x50      & 92 & 24  & ok \nl
\nl
High Resolution Configurations \nl
1984 Dec 05  & A    & 15.0 & 1x25 & 2   & 110 & low S/N; not used \nl
1996 Jan 02  & B    & 15.0 & 2x50 & 2   & 145 & ok \nl
1996 Jan 02  & B    & 8.41 & 2x50 & 3.6 &  36 & low S/N; not used
\nl
1994 May 14  & BnA  & 8.41 & 2x50 & 3.6 &  36 & ok  \nl
1995 Jun 30  & A    & 4.58 & 2x50 & 6   &  71 & ok  \nl
1984 Dec 05  & A    & 4.58 & 1x25 & 6   &  60 & low S/N; not used \nl
 \nl
\enddata 
\end{deluxetable}
\clearpage

\begin{deluxetable}{rrrcrr}
\tablewidth{0pt}
\tablecaption{Henize 2-10 Integrated Radio Flux \label{flux}}
\tablehead{
\colhead{Frequency}  &
\colhead{Flux\tablenotemark{a}}   &
\colhead{Flux\tablenotemark{b}}   &
\colhead{Lum\tablenotemark{c}}   &
\colhead{Flux\tablenotemark{d}}   &
\colhead{Flux\tablenotemark{e}}     \\ [.2ex]
\colhead{(GHz)} &  
\colhead{(mJy)} &
\colhead{(mJy)} &
\colhead{($\times10^{27}erg~s^{-1}~Hz^{-1}$)} &
\colhead{(mJy)} &
\colhead{(mJy)} } 
\startdata
230   &  5.9$\pm$3.3\tablenotemark{f}    &&              
&               &    \nl
96.0  &  6.2$\pm$0.6\tablenotemark{f}    &&              
&               &    \nl
14.9  &  21.1$\pm$1.2   & ...        & 1.9$\pm$0.19  & 17.5$\pm$0.7  &
5.7 \nl
10.6  &  ...            & 40$\pm$15  & ...  & ...    &  \nl 
8.9   &  ...            & 35$\pm$8   & ...  & ...    &  \nl
8.4   &  24.3$\pm$2.4   & ...        & 2.2$\pm$0.2  & 15.6$\pm$0.4  &
4.4 \nl
6.2   &  ...            & 45$\pm$10  & ...  & ...    &  \nl
5.0   &  ...            & 55$\pm$10  & ...  & ...    &  \nl
4.8   &  42.6$\pm$4.3   & ...        & 3.9$\pm$0.4  & 14.5$\pm$0.3  &
3.9 \nl
2.7   &  ...            & 75$\pm$35  & ...  & ...    &  \nl
1.4   &  69.2$\pm$6.9   & ...        & 6.3$\pm$0.6  & ...    &  \nl
0.408 &  ...            & 240$\pm$80 & ...  & ...    &  \nl
0.325 &  100$\pm$20     & ...        & 9.2$\pm$1.8  & ...    &  \nl
\enddata 
\tablenotetext{a}{Data from this paper, using low resolution
VLA configurations and mapping parameters which yield a 
synthesized beam of $\sim$25\arcsec\ FWHM.}
\tablenotetext{b}{From the single-dish  Allen \etal\ 1976 compilation of
        single-dish measurements for comparison}
\tablenotetext{c}{Monochromatic radio luminosity based on a distance of
9 Mpc.
1 Jy = $9.16\times10^{21}~W~Hz^{-1}$ or 
$9.16\times10^{28}~erg~s^{-1}~Hz^{-1}$ at that distance.}
\tablenotetext{d}{Total flux measured in the high resolution maps
        shown in Figure~2 and Figures 3a--b which are made from a restricted
        subset of the visibility data to ensure matched UV coverage.}
\tablenotetext{e}{Contribution from the 5 compact radio knots only,
        as measured by Gaussian deconvolution and recorded in Table~3.
        Uncertainties are $\sim$20\%, and are dominated by the
        variable background levels surrounding the knots.}
\tablenotetext{f}{From the Owens Valley Radio Observatory millimeterwave interferometer
   measurements of Meier \& Turner (1999)}
\end{deluxetable}
\clearpage

\begin{deluxetable}{lrrrrr}
\tablewidth{0pt}
\tablecaption{Properties of Individual Radio Knots \label{flux}}
\tablehead{
\colhead{ }   &
\colhead{Knot 1}   &
\colhead{Knot 2}   &
\colhead{Knot 3}   &
\colhead{Knot 4}   &
\colhead{Knot 5}    } 
\startdata
RA (2000)\tablenotemark{a}  & 08 36 15.014 & 08 36 15.060 & 08 36 15.127 & 08 36 15.234 & 08 36 15.308 \nl
DEC(2000)\tablenotemark{a}  & -26 24 33.81 & -26 24 33.98 & -26 24 34.13 & -26 24 34.00 & -26 24 34.61 \nl 
Peak$_{2~cm}$ (mJy/bm)\tablenotemark{b}      & 0.70$\pm$0.14 & 0.53$\pm$0.10 & 0.56$\pm$0.11 & 1.03$\pm$0.20 & 0.69$\pm$0.14 \nl
Peak$_{6~cm}$ (mJy/bm)\tablenotemark{b}      & 0.82$\pm$0.16 & 0.48$\pm$0.10 & 0.53$\pm$0.11 & 0.62$\pm$0.12 & 0.39$\pm$0.08 \nl
S$_{2~cm}$ (mJy)\tablenotemark{b}            & 0.94$\pm$0.19 & 1.06$\pm$0.21 & 0.89$\pm$0.18 & 1.73$\pm$0.34 & 1.04$\pm$0.21 \nl      
S$_{6~cm}$ (mJy)\tablenotemark{b}            & 0.82$\pm$0.16 & 0.56$\pm$0.11 & 0.89$\pm$0.18 & 1.10$\pm$0.22 & 0.57$\pm$0.12 \nl  
$\alpha^6_2$                                 & 0.11$\pm$0.14 & 0.53$\pm$0.14 & 0.00$\pm$0.14 & 0.38$\pm$0.14 & 0.51$\pm$0.14  \nl        
L$_{2~cm}$ ($x10^{25}~erg~s^{-1}~Hz^{-1}$)\tablenotemark{c} & 8.6$\pm$1.7  & 9.7$\pm$2.0  & 5.2$\pm$1.0 & 9.5$\pm$1.9    & 6.3$\pm$1.2  \nl
L$_{6~cm}$ ($x10^{25}~erg~s^{-1}~Hz^{-1}$)\tablenotemark{c} & 7.5$\pm$1.5  & 5.1$\pm$1.0  & 8.1$\pm$1.6 & 10.1$\pm$2.0   & 5.2$\pm$1.0  \nl
$T_b(2~cm)$\tablenotemark{d}                 & $>$16        & $>$18          & $>$15     & $>$29           & $>$18  \nl
$T_b(6~cm)$\tablenotemark{d}                 & $>$122       & $>$84        & $>$135         & $>$164         & $>$85   \nl
$Q_{Lyc}$ $(10^{51}~photons~s^{-1})$\tablenotemark{e}       & 8.9          & 10.0         & 5.4          & 9.9          & 6.5  \nl
$N_*$ (O7V)\tablenotemark{f}                 & 890          &1000          & 540          & 990          & 650  \nl
$L(H\alpha)$ $x10^{40} erg~s^{-1}$\tablenotemark{g} & 1.2   & 1.4          & 0.7          & 1.4          & 0.9 \nl
\enddata 
\tablenotetext{a}{As measured from a Gaussian fit to the 2 cm map shown
in Figure 3a.}
\tablenotetext{b}{As measured from a 2-D Gaussian fit to the maps shown
in Figure 3a-b.
Approximate 1$\sigma$ uncertainties are 20\%.}
\tablenotetext{c}{For an assumed distance of 9 Mpc.}
\tablenotetext{d}{For the brightness temperature, we use the 
beamsize and peak flux of the maps in Figure~3.
$T_B(K)={{F\lambda^2}\over{2k\Omega}}$ where k is Boltzman's constant,
and $\Omega$ is the beamsize.  In useful units, this becomes
$T_B(K)=1541.5\times4ln2{{F(Jy)\lambda^2(cm)}\over{\pi{a}('')b('')}}$
where $F$ is the flux of the source, $\lambda$ the wavelength of
obsevation,
and $a$ and $b$ are the major and minor FWHM of the source. 
In the absence of information about the actual source structure or size,
we
can only set an upper limit on the actual source by assuming that it
uniformly fills
the syntheized beam, 0.8\arcsec$\times0.40$\arcsec.  This leads to a
lower limit on the brightness temperature.}
\tablenotetext{e}{The number of Lyman continuum photons implied by the
2-cm (15 GHz) luminosity via Equation 2, assuming
that the 2-cm luminosity for each knot is entirely thermal in origin.}
\tablenotetext{f}{The number of O7V stars required to produce the
derived number of Lyman continuum photons,
assuming each star emits 1.0$\times10^{49}~s^{-1}$ (Vacca 1994).}
\tablenotetext{g}{The predicted H$\alpha$ luminosity of each knot based on the
number of Lyman continuum photons derived above.}
\end{deluxetable}
\clearpage